\documentclass[conference]{IEEEtran}
\IEEEoverridecommandlockouts
\usepackage[utf8]{inputenc}
\usepackage{cite}
\usepackage{amsmath,amssymb,amsfonts,amsthm}
\usepackage{algorithmic}
\usepackage{algorithm2e}
\usepackage{hyperref} 
\usepackage{cleveref}
\usepackage{siunitx}
\usepackage{comment}
\usepackage{lipsum} 
\usepackage{xcolor}
\usepackage{graphicx}
\usepackage{acro} 
\let\IDeclareAcronym\DeclareAcronym
\renewcommand{\DeclareAcronym}[2]{%
 \IDeclareAcronym{#1}{%
  #2,foreign-plural={} 
  }
}
\usepackage{caption}
\usepackage{titlecaps} 
\usepackage[version=3]{mhchem}
\usepackage{subcaption}

\def\BibTeX{{\rm B\kern-.05em{\sc i\kern-.025em b}\kern-.08em
    T\kern-.1667em\lower.7ex\hbox{E}\kern-.125emX}}
\begin{document}
\setlength{\parskip}{0cm plus0mm minus0mm}
\renewcommand{\topfraction}{0.85}
\renewcommand{\textfraction}{0.1}
\renewcommand{\floatpagefraction}{0.85}
\setlength{\textfloatsep}{1\baselineskip plus 0.2\baselineskip minus 0.5\baselineskip}

\Addlcwords {a an the at by for in of on to up and as but or nor so yet vs with per} 

\title{\textsc{LoCUS}: \titlecap{A multi-robot loss-tolerant algorithm for surveying volcanic plumes}}

\author{\IEEEauthorblockN{John Ericksen\IEEEauthorrefmark{1}$^1$, Abhinav Aggarwal$^{1}$, G. Matthew Fricke$^{1,2}$ and Melanie E. Moses$^{1,3,4}$}
\IEEEauthorblockA{$^1$Department of Computer Science, $^2$Center for Advanced Research Computing, $^3$ Department of Biology,\\ University of New Mexico, Albuquerque, NM, USA, $^4$Santa Fe Institute, Santa Fe, NM, USA}
Email: \IEEEauthorrefmark{1}johncarl@unm.edu}

\newcommand{\mycomment}[2]{{\color{red} #1: } {\color{blue} #2}}
\newcommand{\abhi}[1]{\mycomment{Abhinav}{#1}}
\newcommand{\john}[1]{\mycomment{John}{#1}}
\newcommand{\paran}[1]{\left( #1 \right)}
\newcommand{\abs}[1]{\left| #1 \right|}
\newcommand{\parfrac}[2]{\paran{\frac{#1}{#2}}}
\newcommand{\floor}[1]{\left \lfloor #1 \right \rfloor}

\newtheorem{definition}{Definition}
\newtheorem{theorem}{Theorem}
\newtheorem{lemma}{Lemma}
\newtheorem{corollary}{Corollary}

\newcommand{\rmax}{\mathsf{R}_{\textsf{max}}}
\newcommand{\rmin}{\mathsf{R}_{\textsf{min}}}

\newcommand{\shortvdots}{\raisebox{3pt}{$\scalebox{.75}{\vdots}$}}





\maketitle

\begin{abstract}
Measurement of volcanic \ce{CO2} flux by a drone swarm poses special challenges. Drones must be able to follow gas concentration gradients while tolerating frequent drone loss. We present the LoCUSalgorithm as a solution to this problem and prove its robustness.  LoCUS relies on swarm coordination and self-healing to solve the task.
As a point of contrast we also implement the MoBSalgorithm, derived from previously published work, which allows drones to solve the task independently.  We compare the effectiveness of these algorithms using drone simulations, and find that LoCUS provides a reliable and efficient solution to the volcano survey problem. Further, the novel data-structures and algorithms underpinning LoCUS have application in other areas of fault-tolerant algorithm research.
\end{abstract}

\acresetall
\section{Introduction}
More than 10\% of the world's population live in the destructive zone of volcanoes, and a quarter of a million people have perished in volcanic eruptions in the last 500 years \cite{Brown2017}. Volcanoes emit unknown amounts of \ce{CO2} and other climate changing gasses, but only 10 of the approximately 300 currently active volcanoes are characterised by long-term datasets that enable any assessment of temporal \ce{CO2} variability \cite{Aiuppa2019}. 
Measuring volcanic \ce{CO2} flux would enable predictions of eruptions, minimizing loss of life and economic impact, as well as informing our understanding of greenhouse gas-driven climate change.

Satellite remote sensing of \ce{CO2} is infeasible, so sampling is currently performed by ground based sensors or aerial surveys with piloted aircraft \cite{Diaz2010}. These techniques are costly, dangerous, and produce temporally and spatially coarse measurements. UAV present an emerging solution \cite{Liu2019} that reduces risk to volcanologists and has the potential to markedly increase sampling resolution within volcano plumes.


An international team of research universities recently demonstrated that UAV can feasibly sample \ce{CO2}  from an active volcano in Papua New Guinea \cite{liu2020}. We developed the dragonfly drone for this task. The dragonfly is capable of measuring \ce{CO2} in real time and has a flight duration of \SI{1}{\hour}.
However, drone loss was very common. Sudden and violent thermal updraughts, acidic plumes, and rugged cliffs were some of the many conditions that destroyed UAV. Further, the remoteness of many survey sites and battery life restrictions necessitate brief missions with small swarms. These hazardous and difficult conditions motivate the need for reliable performance and surveillance algorithms that maximize the chance of completing the \ce{CO2} surveillance task even with the loss of drones, short flight times, and small swarm sizes.



A key task for volcano surveillance is to locate the maximum \ce{CO2} flux (max flux) in a dynamic gas plume. We propose the LoCUS algorithm to maintain a spatially dispersed swarm of drones that can simultaneously measure \ce{CO2} concentrations at different locations and communicate those  measurements across the entire swarm.
We use deductive arguments to prove the loss-tolerance properties of LoCUS, and we test its performance and fault tolerance in simulations.  In particular we show LoCUS guarantees that failed drones are replaced within flight-time proportional to the square root of swarm size, while preserving the swarm symmetry essential to efficient gradient following.  

We hypothesise that maintaining a dispersed team of robots that can simultaneously measure \ce{CO2} at different spatial locations will provide a better estimate of the \ce{CO2} gradient, allowing fast navigation to the \ce{CO2} source. We further hypothesise that the benefit of spatially dispersed measurements outweighs the increased complexity resulting from coordination and self-healing. To test these hypotheses, we develop an alternative approach which allows multiple UAV to independently search for the maximum \ce{CO2} flux. 
We compare LoCUS to MoBS, an algorithm that combines a ballistic search algorithm for multiple agents without communication\cite{aggarwal2019irrational}\cite{Abhinav2019phd} and a gas gradient following algorithm for robots inspired by Moth pheromone tracking\cite{Li2006}\cite{belanger1998biologically}. 
 



\begin{figure}[t]
    \centering
    \includegraphics[width=\columnwidth]{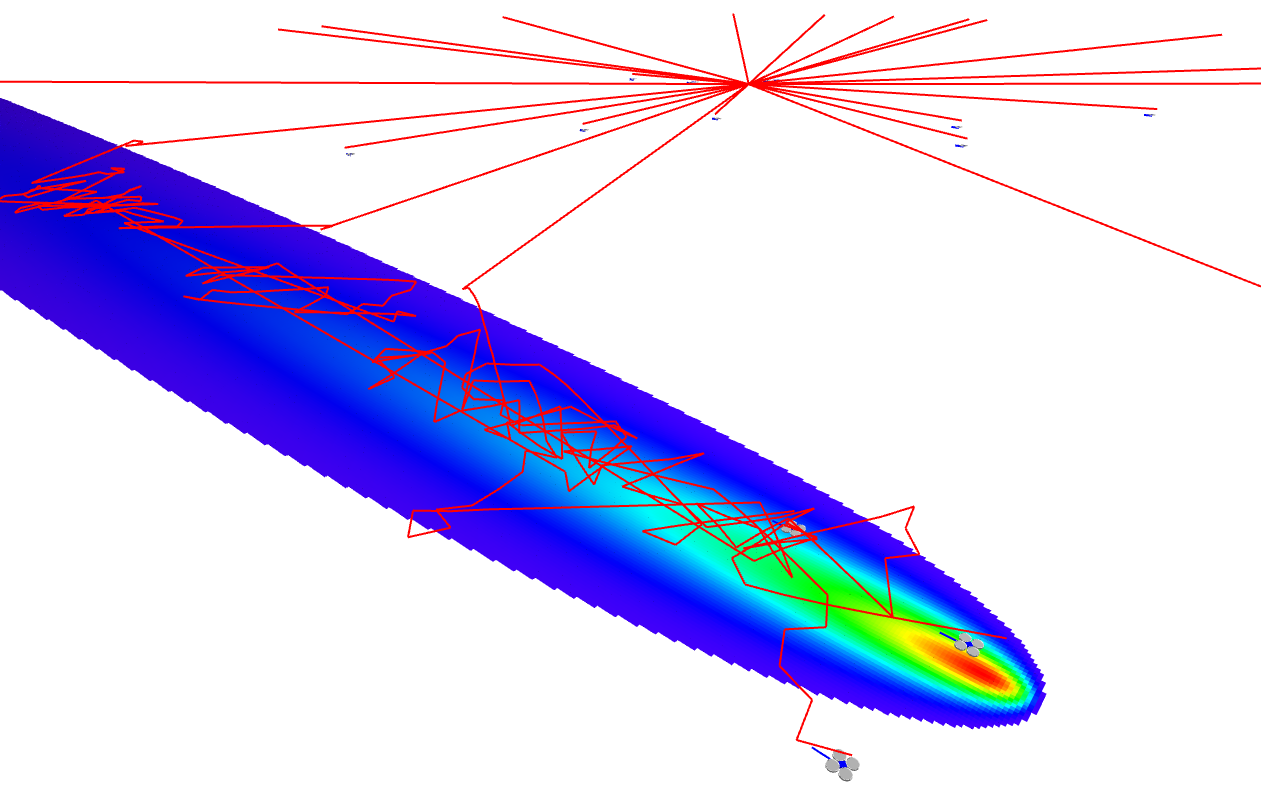}
	\caption{MoBS simulation with 16 drones and a smooth plume.  The red lines trace each drone's independent search for the plume using golden ratio spokes from the center of the arena. After each drone contacts the plume, it switches to a Moth pheromone inspired search algorithm to find the max flux.}
	\label{fig:mobsScreenshot}
\end{figure}

\section{Related Work}

An algorithm for reliably locating max flux using a remote-sampling robotic-platform requires the following:
\begin{enumerate}
  \item \emph{Search:} A search pattern to explore an area to make initial contact with the plume.
  \item \emph{Plume Gradient Following:} After plume contact is made, the platform follows the gas plume to the source.
  \item \emph{Failure Resistance:} The collection of robots needs to respond to failures to maintain a cohesive structure.
\end{enumerate}

Schleich et. al. \cite{Schleich2013} proposes \emph{searching} an area using a fully-connected swarm of drones and compares this against a random and pheromone-following approach.  They find that a fully-connected swarm satisfies base-station connectivity requirements while achieving slightly better survey performance for larger swarm sizes.  This motivates LoCUS, as keeping the swarm in contact provides benefits that outweigh the overhead of maintaining swarm connectivity.

Neumann et. al. \cite{Neumann2015} compares 3 algorithms for \emph{plume gradient following}: the surge-cast algorithm, the Dung Beetle (zig-zag) algorithm, and the pseudo-gradient algorithm using a single robot agent. Through the author's experiments, in both simulation and physical robots, they validate all three algorithms promising for micro UAVs each under different ciricumstances.  Our approach uses multiple robots for plume gradient following, with MoBS closely resembling the surge-cast algorithm and LoCUS resempling pseudo-gradient algorithm across the swarm formation.

Chen et. al. \cite{chen2017using} apply a Particle Swarm Optimization algorithm to follow a gas plume gradient in an indoor environment. This approach requires full swarm connectivity to communicate global arena information throughout the swarm. This motivates keeping the swarm connected with coordinated movement for gradient descent.

In \cite{Cabrita2014}, Cabrita et. al. investigate locating the max flux using Gaussian parameter estimation leveraging a simulated annealing error minimisation approach. They test this algorithm successfully on a swarm of 5 robots. We implement a similar model in MoBS and LoCUS, but we only use the local gradient to navigate the plume in the case of MoBS, and the gradient that spans the swarm's full extent in LoCUS. We use their simple linear fit to determine the direction of the \ce{CO2} gradient.

\begin{figure}[t]
    \includegraphics[width=\columnwidth]{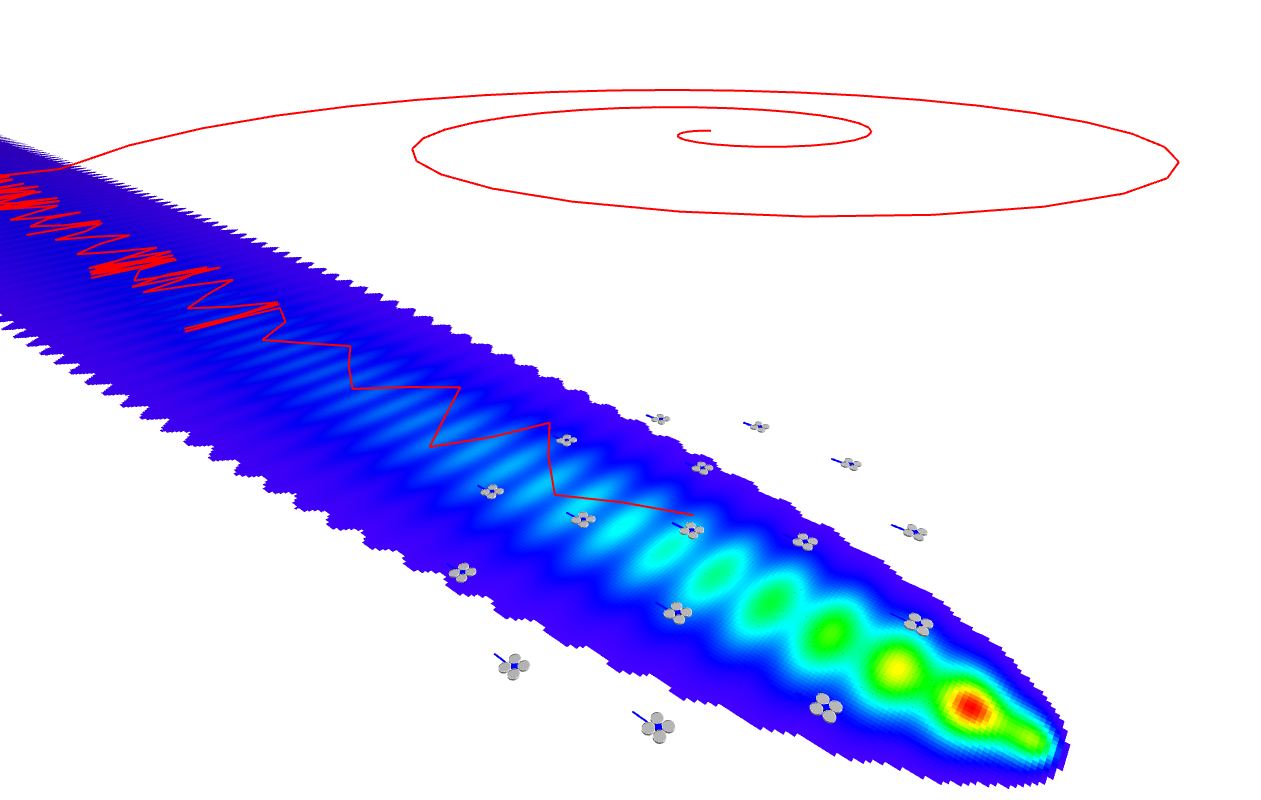}
	\caption{LoCUS simulation with 16 drones and a perturbed plume.  The red lines trace the swarm's Archimedes Spiral search for the plume. After contacting the plume, the swarm follows leverages its simultaneous spatially dispersed measurements to descend the gradient to the max flux.}
	\label{fig:locusScreenshot}
\end{figure}

Flocking algorithms are effective at coordinating movement while being \emph{failure resistant}.  Souissi et. al. \cite{souissi2011oracle} and Yang et. al. \cite{yang2011fault} propose leader based approaches for moving a swarm flock while maintaining a given shape and detecting and recovering from failures. Their algorithms keep the swarm together during movement. LoCUS, on the other hand, makes theoretical guarantees about swarm symmetry as drones are lost, given a small collection of drones in close enough proximity that all drones can maintain communication with each other.  Our approach could be applied to heal traditional flocking algorithms like the one presented in \cite{viragh2014flocking}.


Paliotta et. al. \cite{paliotta2015adaptive} present a \emph{plume gradient following} agent based model for three fully networked agents \cite{breivik2008ship}\cite{biyik2007gradient}\cite{bachmayer2002vehicle}.  We extend this structured plume gradient following approach with LoCUS by increasing the swarm size, tuning agent capabilities to mimic our dragonfly robotic platform, and adding a fault recovery mechanism.

\section{Derivation and Analysis of LoCUS}

The LoCUS algorithm ensures a fully connected swarm with efficient recovery from drone failures.  A LoCUS swarm is able to be controlled as a single unit, by directing all members of the swarm at once.  
We first discuss the basic algorithm assuming no failures, 
and then discuss how the swarm recovers from drone failures and resumes its mission. 

Let $N$ be the total number of autonomous drones in the system. Each drone has a unique ID in $\{1,\dots,N\}$ and a communication radius $\rmax$ and a safety radius $\rmin$. Each drone can communicate with any drone within $\rmax$ distance, but requires a minimum distance between any two drones in the swarm to be $\rmin$ to avoid collisions.

\subsection{Balanced Range-Limited Trees}\label{sec:tree}

\begin{figure}[t]
	\centering
	\includegraphics[width=\columnwidth, trim={0cm 1.5cm 12cm 3.7cm}, clip=true]{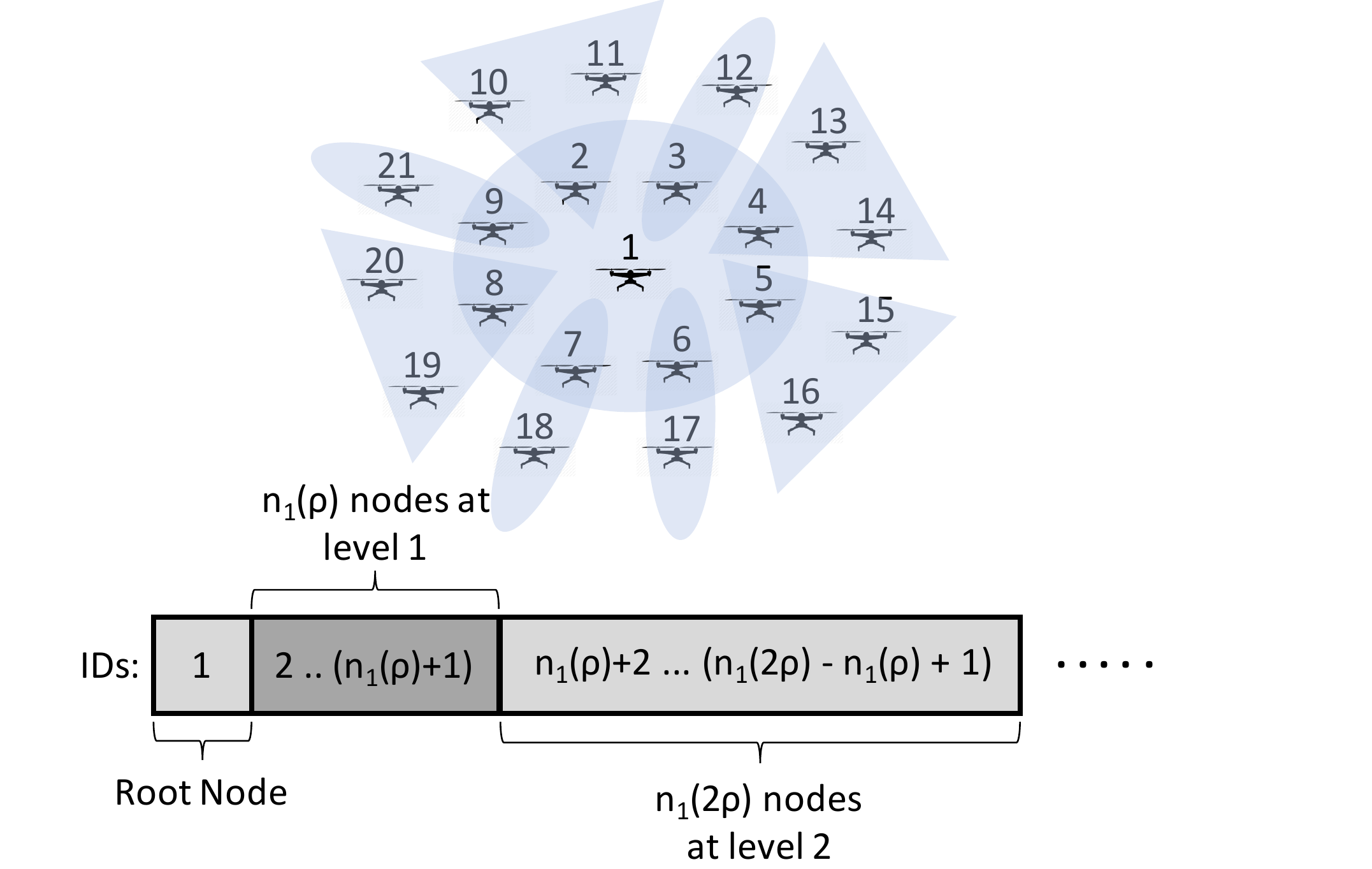}
	\caption{Assignment of nodes to levels based on their IDs. Since the number of nodes at each level is fixed, the assignment is deterministic and can be computed locally to determine placement in the swarm (see \Cref{sec:formation}).  The blue regions denote parent/child communication links.}
	\label{fig:levelFormation}
\end{figure}


\begin{figure*}[t]
	\centering
	\includegraphics[width=\textwidth, trim={0cm 7.5cm 1cm 4cm}, clip=true]{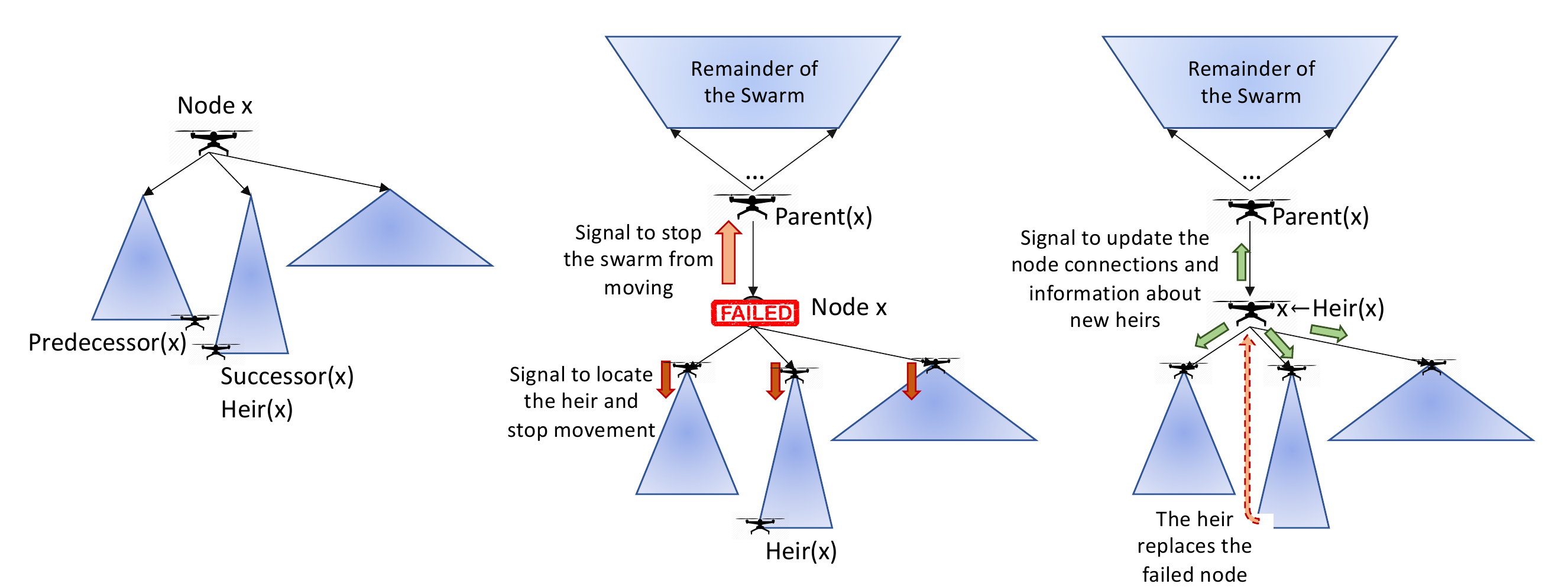}
	\caption{A schematic of single failure recovery in the LoCUS algorithm. When a node fails, a signal is sent to its parent and children to stop the swarm movement and inform the heir. The heir node then travels to the location of the failed node and the neighboring nodes update their local information.}
	\label{fig:singleFailureSchematic}
\end{figure*}

\begin{definition}
	Given $R_{min},R_{max} > 0$ and an integer $n > 0$, an $(R_{min},R_{max})$-Range-Limited Tree on $n$ nodes is a rooted tree, where the distance between any two nodes is at least $R_{min}$ and at most $s$. In particular, a maximal $(R_{min},R_{max})$-Range-Limited Tree is one in which the distance between the parent node and any of its children is $R_{max}$. The ratio $\rho = R_{max}/R_{min}$ is the \emph{spread} of this tree.
\end{definition}

As with standard $k$-ary trees, we can define the height of a Range-Limited Tree $\mathcal{T}$ node in terms of the heights of its children. We define the height of the root node as zero and then, recursively, the height of $\mathcal{T}$, denoted $\textsf{height}\paran{\mathcal{T}}$, as $\textsf{height}\paran{\mathcal{T}} = 1 + \max_{i} \left\{ \textsf{height}\paran{\mathcal{T}_i} \right\}$, where the maximum is over the height of all children $\mathcal{T}_i$ of $\mathcal{T}$. Similarly, we define the level of a node as $\textsf{level}(\mathcal{T}_i) = 1 + \textsf{level}(\mathcal{T}_i.parent)$, where $\mathcal{T}_i.parent$ is the parent node of $\mathcal{T}_i$. For this recurrence, the root node is defined to be level zero. Thus, the root node has the largest height in the tree but is located at the lowest level.

\begin{definition}
	Let $\mathcal{T}$ be a Range-Limited Tree. We say that $\mathcal{T}$ is \emph{Balanced} if for every node in $\mathcal{T}$, the difference in the heights between any two of its children is at most one, i.e., for every node $\mathcal{T}_i \in \mathcal{T}$ with children $\mathcal{T}_i^{(1)},\dots,\mathcal{T}_i^{(m)}$, it must hold that $\abs{\textsf{height}\paran{\mathcal{T}_i} - \textsf{height}\paran{\mathcal{T}_j}} \leq 1$ for all $i \neq j$. 
\end{definition}

Each node maintains a pointer to its \emph{heir} in the tree. This is crucial to achieve fault tolerance in LoCUS. We define the heir of a node as its \emph{successor}, if it exists, or its \emph{predecessor}, otherwise.  If neither a successor or predecessor exists, the node is a leaf node and the heir is null.  To define a successor node, we first define an \emph{inorder} traversal of the tree, denoted $\textsf{in}(\mathcal{T})$. Let $\mathcal{T}^{(1)},\dots,\mathcal{T}^{(m)}$ be the children of the root node for $\mathcal{T}$. Then, the inorder traversal of $\mathcal{T}$ prints the IDs of these nodes in the following order (here, $\cdot$ represents the concatenation operator): $\textsf{in}(\mathcal{T}^{(1)}) \cdots \textsf{in}(\mathcal{T}^{\floor{\frac{m}{2}}}) \ \cdot \ \textsc{ID}(\mathcal{T}) \ \cdot \ \textsf{in}(\mathcal{T}^{\floor{\frac{m}{2}}+1}) \cdots \textsf{in}(\mathcal{T}^{(m)})$. Note that the inorder traversal is unique for a given tree. We can now define the successor and predecessor of a node. 

\begin{definition}
	Node $\mathcal{T}_j$ is a successor of the node $\mathcal{T}_i$ in the tree $\mathcal{T}$ if $\textsf{ID}(\mathcal{T}_j)$ immediately follows $\textsf{ID}(\mathcal{T}_i)$ in the inorder traversal of $\mathcal{T}$. Similarly, we say that $\mathcal{T}_j$ is a predecessor of $\mathcal{T}_i$ if $\textsf{ID}(\mathcal{T}_j)$ immediately comes before $\textsf{ID}(\mathcal{T}_i)$ in the inorder traversal of $\mathcal{T}$.  In all cases, a node is either a leaf, or either its successor or predecessor is a leaf node of tree $\mathcal{T}$.
\end{definition}

\subsubsection{Formation Algorithm}\label{sec:formation}
The LoCUS algorithm swarm takes the shape of an $(\rmin,\rmax)$-Balanced Range-Limited Tree. A balanced Range-Limited Tree layout obtains maximal spatial coverage while maintaining a minimum separation between drones to avoid collisions, and keeps drones within communication range.

\begin{lemma}[Number of Nodes at a Given Level]
\label{lem:numNodesLevel}
	Let $\mathcal{T}$ be a maximal balanced $(r,s)$-Range-Limited Tree on $N$ nodes with $\rho = s/r$. Then, the number of nodes at level zero is given as $n_0(\rho) = 1$, and for each $k > 0$, the number of nodes at level $k$ is $n_k(\rho)$ = $\floor{\frac{2\pi}{\sin^{-1}\paran{\frac{k}{\rho}}}}$.  This is the calculation of the whole number of nodes that fit on a circle at radius $s \times k$ separated by distance $r$.
\end{lemma}

Drones deterministically compute their location in the swarm with respect to tree layout. This computation is local to the drones and can be calculated purely by the drone IDs (see \Cref{fig:levelFormation}). From \Cref{lem:numNodesLevel}, we know that the number of drones at level $k$ is $n_k(\rho)$. Thus, the space of drone IDs can be partitioned based the levels in which the drones belong. For example, the drone with ID 1 is the root node and has level zero, whereas the drones with IDs from $2$ to $n_1(\rho)+1$ all belong to level one. 

Each node (besides the root) in the LoCUS tree structure holds a parent reference and list of children.  This facilitates bidirectional communication throughout the swarm, as required by the LoCUS algorithm.  Parent nodes are calculated by the closest node in the previous layer.

\begin{lemma}[Number of Levels]\label{lem:numLevels}
	The number of levels in a maximal balanced $(r,s)$-Range-Limited Tree on $N$ nodes with $\rho = s/r$ is $O\paran{\sqrt{N/\rho}}$.
\end{lemma}
This is also a bound on the maximum height of the tree and hence, the maximum number of communication hops required for any node in the tree to transmit a message to any other node in the tree. In particular, when $\rho$ is low (i.e. when the communication radius is not too large compared to the safety radius), then the diameter of the tree is $O(\sqrt{N})$, however, when the communication radius is large, say with $\rho = \Omega\paran{\frac{N}{\log N}}$, then the diameter of this tree becomes $O(\log N)$, which is similar to that of a tree with constant arity. Communication is highly efficient in this case and the latency for transmitting messages is low. 

\subsubsection{Insertion of Nodes}\label{sec:insertion}
Always insert at the first available leaf node so that insertion cost is $O(1)$. Insertions do not affect the balance of the tree, since no new levels are created unless the previous level is completely full.

\subsubsection{Deletion of Nodes}
Replace the deleted node by its heir. If the deleted node is a leaf node, then there is no heir replacement for this node and hence, there is no deletion cost. However, when a node at height $h \geq 1$ fails, then its heir is located at a communication hop distance of $O\paran{\sqrt{N/\rho} - h}$ from this failed node. Hence, although only $O(1)$ link changes happen upon this replacement, the total number of messages sent is $O\paran{\sqrt{N/\rho} - h}$.

\subsection{Handling Drone Failures}\label{sec:handlingFailures}
The use of the Balanced Range Limited Trees data structure offers the swarm resilience against arbitrarily many crash failures, even when all but one drone remains in the system. We achieve this robustness as follows (see \Cref{fig:singleFailureSchematic}) -- When a node fails, a signal is sent to its parent and children to stop the swarm movement and inform the heir. This signal can be sent out when the node believes it is about to fail, for example when its battery is critically low, or by its parent and children when it fails to respond to a heartbeat.  Upon receiving this signal, the heir node travels to the location of the failed node and replaces it in the swarm.  Finally, because the tree structure changed, heirs are recalculated on ancestor nodes of the replacement heir node's original leaf location.
 
Since each drone stores its heir information, it can directly inform the heir drone when it believes a failure is inevitable. For the case when the failure happens without any signal being sent out, the child drones use their \emph{parentHeir} field to contact the heir drone in the swarm for recovery. If there are no child drones, then the failed drone does not require any recovery mechanism since it is already in the last layer of the swarm.

\begin{figure}
	\centering
	\includegraphics[width=0.7\textwidth, trim={7.8cm 13.5cm 2cm 6cm}, clip=true]{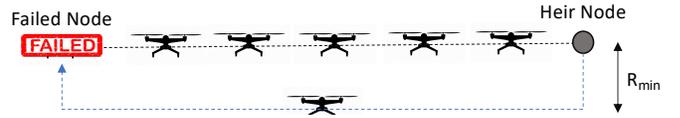}
	\caption{The heir replaces the failed node by flying under the swarm at a safe distance to prevent collisions.}
	\label{fig:droneMovementUnder}
\end{figure}

To ensure that drones do not collide with other drones while the swarm is rearranged, the heir drone descends to a distance of $\rmin$ and travels at this height to its destination (see \Cref{fig:droneMovementUnder}), at which point it climbs back up to the given elevation. The swarm must stop moving during this recovery phase to avoid complicating communications and movement when the swarm is disconnected.

This heir-based recovery scheme achieves a reformation cost of $O\paran{\sqrt{N/\rho}}$ -- the bound on $\textsf{height}\paran{\mathcal{T}}$ -- by only inducing local adjustment in the swarm, without disturbing other drones.  Note that moving one drone to replace its heir would be approximately equal to drones between the heir and the failing drone shifting up in the tree.

\subsection{Handling Simultaneous Drone Failures\label{sec:simultanious_failures}}
If both a drone and its heir drone fail at the same time, and the swarm uses the algorithm above to simultaneously recover from both, then it will enter a deadlock scenario. We introduce the following algorithm for handling simultaneous failures with the caveat that it requires global knowledge of the swarm state to execute.  A more advanced distributed version of this algorithm that executes without global knowledge is possible, but we leave this analysis and implementation for future work.


\emph{Outer-Level First (OLF):} In this scheme, we use the fact the failures in outer levels of the swarm cost less to recover than failures on the inner levels.  This is because the distance to the heir node is smaller in outer levels.  For example, leaf nodes may be removed from the swarm outright without replacement, a node at $\textsf{height}\paran{\mathcal{T}}/2$ would require its heir to move $\rmax\:\textsf{height}\paran{\mathcal{T}}/2$ distance, and the root node would require its heir to move $\rmax\:\textsf{height}\paran{\mathcal{T}}$ distance to replace. Thus, whenever a node gets a failure signal, it first checks to see if there is any existing failure recovery that is active in any of its children. If yes, it waits for those to finish, then proceeds to process the signal from its parent.  Concretely, this is implemented by gathering a set of failures across the swarm, and processing them in descending order by $\textsf{height}\paran{\mathcal{T}_i}$.

\section{The MoBS Algorithm and its Implementation}

The MoBS algorithm takes a different approach to the max flux problem by allowing each UAV to navigate independently. Each  UAV starts at the center of the arena and picks a uniformly random angle between \SI{0}{\degree} and \SI{360}{\degree} and sets 100 waypoints in \SI{1}{\meter} increments from the center in that direction to produce spokes to search the arena.  At each waypoint the UAV collects a gas plume sample and reacts accordingly.  The UAV continues to follow the spoke waypoints if a reading of less than 0.005. Otherwise, the UAV changes strategies into the moth-pheromone chemotaxis algorithm inspired by \cite{Li2006}.  Subsequent spokes are build by adding $2\pi/\phi$ \si{\radian} to the previous spoke angle where $\phi$ is the golden ratio 1.618 that has been shown to search best given no communication amongst members of the swarm \cite{aggarwal2019irrational}.

The moth-pheromone chemotaxis algorithm compares the gas reading at the current time step against the previous time step and determines if the signal has increased or decreased.  If the signal increased then the drone continues moving in the same direction.  If the signal stays the same or decreases then the drone moves in a new uniformly random direction.  A zero signal detected for greater than 4 time steps reverts the drone back to continue the golden ratio driven spoke search algorithm. Because there is never any communication among UAV in MoBS, failed UAV stop collecting samples but have no impact on other UAV. 

\section{Experimental Methods}

 We measure performance of both algorithms for a range of swarm sizes and failure scenarios in simulation. Given the practical limitation of battery life on flight time, our primarily interest is minimizing the time to find the max \ce{CO2} flux. We halt the simulation when the max flux is found (a drone samples within \SI{1}{\meter} of the max flux location), if the entire swarm is in a failed state, or when \SI{17.3}{\hour} of simulation time has passed ($10^6$ time steps).

We implement the LoCUS and MoBS algorithms in Autonomous Robots Go Swarming (ARGoS) \cite{pinciroli2011argos}.\footnote{Implementation source code can be found at \url{https://tinyurl.com/tne7tzu}}  ARGoS is a C++ and Lua based physics multi-robot simulator and is suitable for proof-of-concept simulations, while preserving realistic physical dynamics with the DYN3D physics engine. We use ARGoS to simulate Spiri UAV (Pleiades Robotics Inc) including 3 dimensional locality (GPS) inputs and go-to coordinate capabilities. Additionally, we are able to command $N$ drones in the simulation.  These capabilities make ARGoS a natural fit for experimental investigation of LoCUS and MoBS.

The gas plume is modeled in ARGoS as a simple two dimensional slice of a Gaussian plume \cite{stockie2011mathematics} with a source max flux location ($x$ and $y$), stack height ($H = \SI{10}{\meter}$), wind speed ($u = \SI[per-mode=symbol]{50}{\meter\per\second}$), emission rate ($Q = \SI[per-mode=symbol]{2}{\kilogram\per\second}$), and diffusion rate ($K=\SI[per-mode=symbol]{1}{\kilogram\per\second}$):

\begin{equation} \label{eq:simpleplumeequation}
\textsc{unperturbed}(x, y) = \frac{Q}{2 \pi K x} \: \exp\paran{- \frac{u  (y^2+H^2)}{4  K x}}
\end{equation}

The source of the plume is located at a uniformly random location in the simulation within \SI{100}{\meter} of the UAV take-off location.  Each UAV may detect the gas concentration at its given coordinate as a floating point value between 0 (low) and 1 (high) gas concentration signals.  To limit the experimental variance, we only vary the location of the plume and not the shape, intensity, or rotation of the plume.  We test the algorithms against the smooth plume described in (\ref{eq:simpleplumeequation}) and a perturbed version of the plume designed to make following the gradient more realistic and challenging:

\begin{equation} \label{eq:perturbedplumeequation}
\textsc{perturbed}(x, y) = (0.8 + 0.2 \sin(4x)) \: \textsc{unperturbed}(x, y) 
\end{equation}

Our two failure models in these experiments are motivated by flying a swarm of UAV to gather volcano gas \ce{CO2} emission data.

\emph{Generic Failures:} To represent a UAV battery failure, crashes, or other miscellaneous failures that increase in likelihood as flight time increases, we use a uniform failure probability per drone per time step given by $p_f > 0$, which depends on the number of drones existing in the system at time $t$.

\emph{In-Plume Failures:} We use the gas plume emissions reading $r$ at time $t$ to drive the probability of failure on each drone given by $p_f r > 0$.  This models the higher probability of failure as corrosive gases or temperatures associated with more concentrated volcano gas emissions are encountered.

Drone failure is represented by a boolean flag on the drone controller that, if enabled, stops the drone from moving or receiving further waypoints from its parent.  Once a drone fails, it is never recovered.

\subsection{Implementation of LoCUS}

The LoCUS algorithm arranges members of the swarm by distributing each drone through space using specified $\rmin$ and $\rmax$.  The unique ID of each member of the swarm allows a unique 2D location offset from the central root node to be calculated. The drones are distributed in a plane by each offset using a constant height of \SI{10}{\meter}.  Each drone's parent is assigned by finding the closest drone in the previous shell of the swarm.  This parent/child relationship constructs the data structure pivotal to maintaining communication throughout the swarm.

We implement a recursive algorithm to distribute navigation waypoints by communicating them from the root drone down through its children, to its children's children, and so on.  When waypoints are distributed, the swarm offset location for each drone is added to the waypoint to ensure that the swarm maintains $\rmin$ and $\rmax$. 

At takeoff, the root LoCUS drone is given the initial starting position.  Using the recursive waypoint distribution, this initial starting position waypoint directs the swarm to assemble the shell structure exhibited by \Cref{fig:levelFormation}.

To make initial contact with the plume, the swarm is directed from the root to follow the Archimedes' spiral.  For coordinated swarm search, the Archimedes' spiral has been shown to find targets faster than a spoke algorithm \cite{aggarwal2019ignorance}.  This search pattern is created by building waypoints along the spiral.  Each waypoint is calculated to space the arms of the spiral by the radius of the largest full swarm shell and an incremented angle.  Using the radius of the swarm  ensures that we have full coverage of the simulation arena.

After a waypoint is reached, a plume gas reading is sampled from each drone and communicated via the tree structure up to the root drone where the readings (\emph{val}) and associated gps coordinates (\emph{x}, \emph{y}) are aggregated into the \emph{uav} array. The aggregated data is input into matrix and vector form and fit with a slope (\emph{b}) using linear regression in the form $Ab+\epsilon=y$ by minimizing $\epsilon$ through least squares approximation provided by the Eigen C++ library \cite{jacob2012eigen}:

\begin{equation} \label{eq:linearRegression}
\underbrace{\begin{bmatrix}
        1 & uav[1].x & uav[1].y \\
        1 & uav[2].x & uav[2].y \\   
         \shortvdots & \shortvdots & \shortvdots
    \end{bmatrix}}_{A}
    \underbrace{\begin{bmatrix}
        b[0] \\
        b[1] \\
        b[2]
    \end{bmatrix}}_{b}
    +
    \underbrace{\begin{bmatrix}
        \epsilon[0] \\
        \epsilon[1] \\
        \shortvdots
    \end{bmatrix}}_{\epsilon}
    =
    \underbrace{\begin{bmatrix}
        uav[1].val \\
        uav[2].val \\
        \shortvdots
    \end{bmatrix}}_{y}
\end{equation}

The slope of this linear fit ($b[1], b[2]$) is used to provide a normal vector to direct the swarm to perform a gradient descent in the direction of the highest plume signal.  If a zero magnitude linear slope is found, then the swarm continues to follow the Archimedes spiral.

Failures are handled as follows. First, once a waypoint is reached, failures in the swarm are queried for from the root.  This is a recursive call, similar to the waypoint distribution, to gather a set of the failed status of the entire connected swarm. For these experiments and for simplicity, we implement the Outer-Level First (OLF) scheme failure recovery model.  This scheme requires global knowledge of the swarm as it uses failed drones to determine the status of their children.  We then proceed to heal the swarm as outlined in \ref{sec:handlingFailures} Handling Failures remove each of the failed drones and replace them by their heir in order, waiting for each heir to take the place of the failed drone before proceeding to the next failed drone.  After the replacement, heirs of all ancestors up to the root are recalculated to take into account the change in the swarm. Of course, if a failed drone is found to have no heir (a leaf drone) then they are removed from the swarm without replacement.

Once all failures are processed a swarm re-balance is executed to ensure a consistent minimum radius to the swarm.  This iteratively removes leaf drones from the deepest branches of the tree and inserts them into the shallowest branches of the tree.  The root node executes this operation until $\textrm{height}_{max}-\textrm{height}_{min} \leq 1$.  Once there are no more failures in the swarm and the swarm is re-balanced, then the next waypoint is calculated and the swarm movement continues.

We observed corner-case scenarios where the swarm oscillated between two points, never moving towards the max flux location. To resolve this, we added both a \SI{0.1}{\meter} random offset and a uniformly random rotation to the swarm between \SI{0}{\degree} and \SI{45}{\degree} at each waypoint. This randomization strategy allowed to swarm to exit these oscillating corner cases.

\subsection{Experimental Setup}

The experimental factors we explore in simulation are  swarm size, whether the plume gradient is smooth or perturbed, the failure rate, and whether the failure rate increases with gas concentration.  For the LoCUS algorithm, we set $\rmin =$ \SI{3}{\meter} and $\rmax =$ \SI{3}{\meter}.  The response variables are whether the max flux was found, the elapsed time before encountering the plume, and the total time taken to find the max flux.

To compare LoCUS and MoBS we perform the following four experiments.  In experiment 1, we compare the time to find max flux of the smooth plume for LoCUS and MoBS in 100 trials with both failure probabilities set to 0 by varying the swarm size from 2 to 20 UAV.  In experiment 2, we duplicate experiment 1 using the perturbed plume.  These first two experiments are designed to compare the times to encounter the plume and navigate to the maximum flux of LoCUS and MoBS without failures.  In experiment 3, we vary the generic failure probability from $10^{-1}$ to $10^{-6}$, set the in-plume failure probability to 0 (so that the probability of failure is the same inside and outside of the plume), and use the smooth plume over 100 trials with a swarm size of 20.  In experiment 4, we duplicate experiment 3 but vary the in-plume failure probability from $10^{-1}$ to $10^{-6}$ and set the generic failure probability to 0.  The last two experiments measure the impact of failures on LoCUS and MoBS. In each of these experiments we compare the performance of LoCUS with healing enabled and disabled. This enables us to assess whether maintaining symmetry though healing is worth the time taken to repair the swarm. 

\section{Experimental Results}

\begin{figure}
	\centering
	\includegraphics[width=\columnwidth, trim={0cm 0cm 1.5cm 0cm}, clip=true]{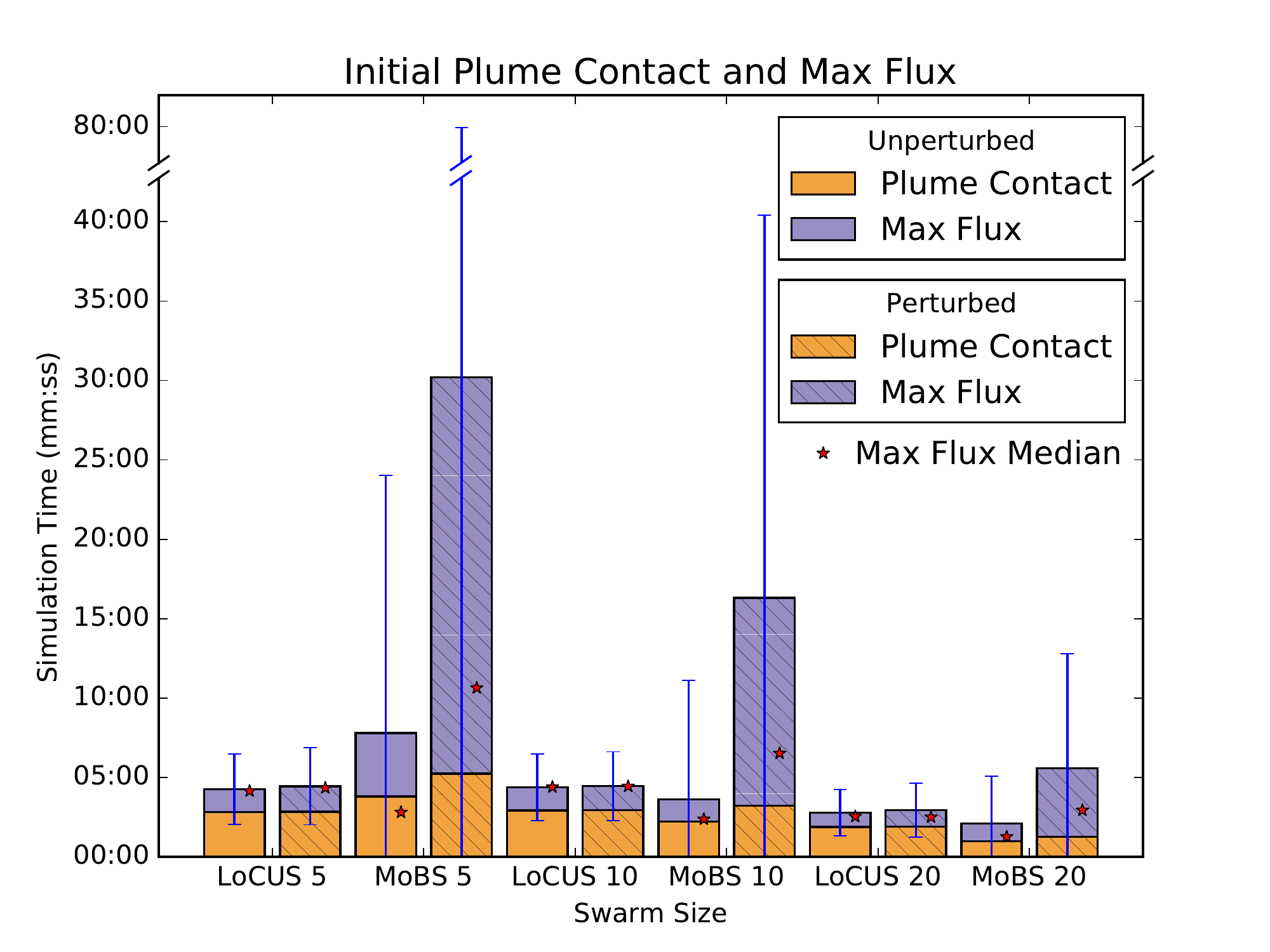}
	\caption{Time to find the max flux (top, purple) and initial plume contact (bottom, orange) for LoCUS and MoBS for swarm sizes 5, 10, and 20 in the unperturbed and perturbed plumes. Stars show the median time to max flux and error bars are one standard deviation centered at the mean. 
	}
	\label{fig:perturbedPlumeEncountered}
\end{figure}

\subsection{Experiment 1: Unperturbed Navigation}

Experiment 1 compares the time to reach max flux of the unperturbed in swarms of 5 to 20 UAV for LoCUS and MoBS depicted in Figure \ref{fig:perturbedPlumeEncountered} with the unhashed bars.  We find that, for smaller swarm sizes (up to the 5 UAV shown), the average time and standard deviation to plume contact and max flux for LoCUS is significantly smaller than MoBS.  For larger swarm sizes (10, 20), MoBS reaches plume contact on average faster. LoCUS navigates from plume contact to max flux in about the same time as MoBS, but LoCUS has less variance in time to achieve both plume contact and max flux.



\subsection{Experiment 2: Perturbed Navigation}

Experiment 2 extends experiment 1 using the perturbed plume depicted in Figure \ref{fig:perturbedPlumeEncountered} with the hashed bars.  We find that the difference in plume dynamics significantly increases the average time to max flux for the MoBS algorithm, but the LoCUS time to max flux remains short.  This increases the average time to max flux for MoBS so that is slower than LoCUS 
for all swarm sizes tested.  Additionally, the standard deviation for MoBS is much larger than that of LoCUS which is partly driven by several outliers that lasted up to 344, 124, and 31 minutes for 5, 10 and 20 UAV respectively. These times risk failure for the drones to return given the UAV battery capacity.





\subsection{Experiment 3: Generic Failure Effects}
\begin{figure}
	\centering
	\includegraphics[width=\columnwidth]{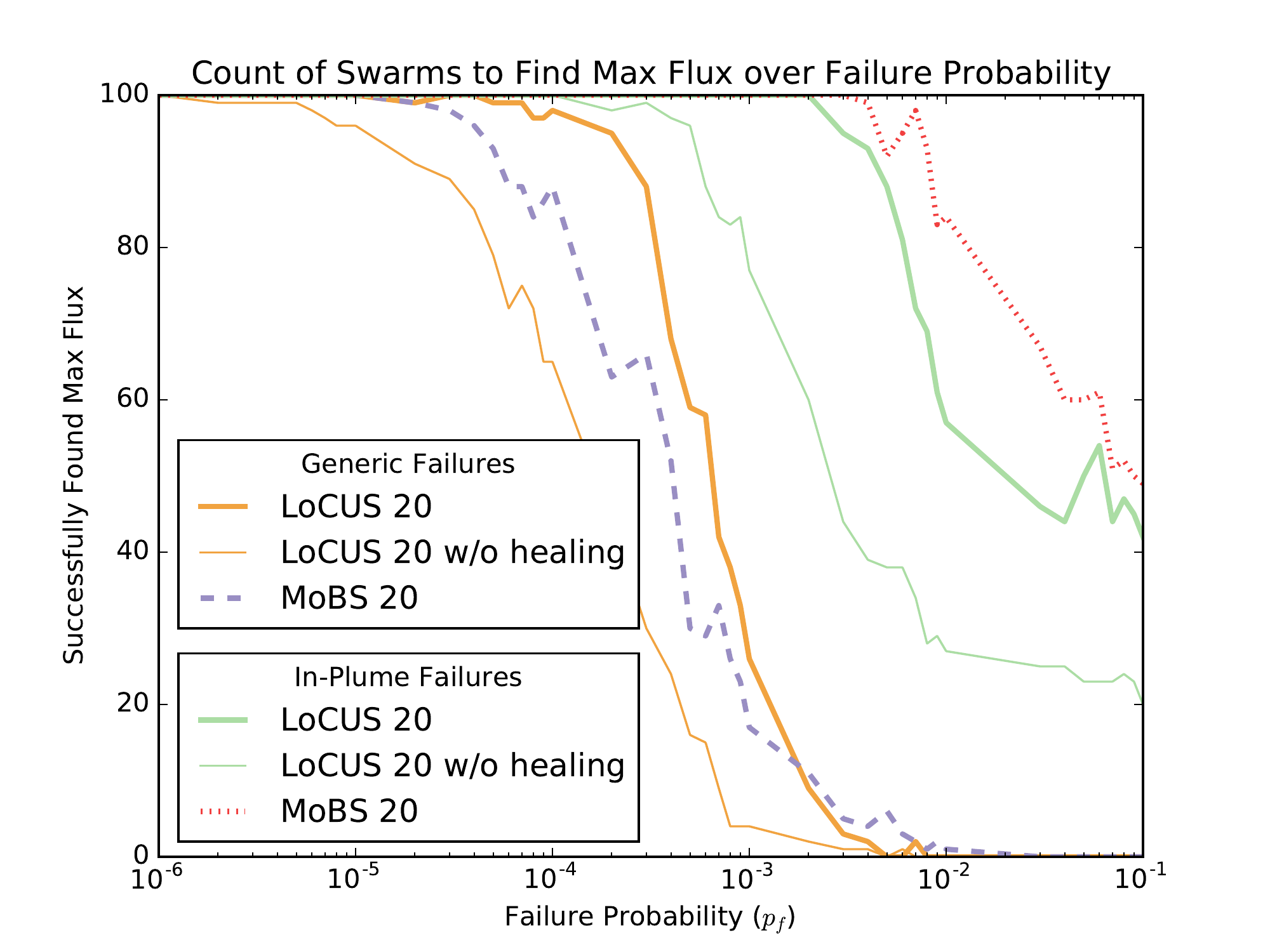}
	\caption{Success rates in 100 trials with generic and in-plume failures with 20 drones.  The left thin solid lines are LoCUS with healing disabled and the right thick solid lines are LoCUS with healing enabled.  The two left orange thin and thick solid lines are LoCUS with generic failures, while the green right solid lines are LoCUS with in-plume failures.  MoBS success counts are graphed using dashed and dotted lines -- the left dashed line is MoBS with generic failures and the right dotted line is MoBS with in-plume failures.  This shows how significant healing is to LoCUS successfully completing the max flux location task.}
	\label{fig:genericFailureMaxFlux}
\end{figure}

Experiment 3 
includes generic failure probabilities from $10^{-1}$ to $10^{-6}$ including a specialized version of LoCUS that does not heal from failures as depicted in Figure \ref{fig:genericFailureMaxFlux}.  We find that, with generic failures, LoCUS and MoBS both respond similarly to the failure probability by beginning to unsuccessfully complete the max flux location task between the probability of failures of $10^{-4}$ and $10^{-3}$.  This is contrasted against the LoCUS without healing that responds much earlier to the probability of failure at about $10^{-5}$.

\subsection{Experiment 4: In-Plume Failure Effects}

Experiment 4 extends experiment 3 using the in-plume failure model. With in-plume failures, LoCUS and MoBS both begin to fail to complete the max flux location task at failure rates of $10^{-2}$.  This is contrasted against the LoCUS without healing that fails to complete the task with much lower UAV failure probabilities of about $10^{-4}$.


\section{Discussion}

LoCUS provides a failure-tolerant structure for exploring and pinpointing the max flux location of a \ce{CO2} plume. LoCUS guarantees that a group of drones can communicate to each other simultaneous spatially dispersed measurements which can be used to calculate a gas gradient better than an individual drone. This is particularly useful for finding the location of maximum flux in a perturbed plume such as those produced by volcanos in dynamic environments. LoCUS provides a way re-form the swarm given the inevitable failure of drones in hazardous conditions present when monitoring gas efflux from volcanoes. We compare LoCUS with the fully dispersed MoBS algorithm and show in experiments 1 and 2 that the LoCUS algorithm is able to find the max flux of a plume, both smooth and perturbed,  at least as fast as the MoBS algorithm in expectation, but with substantially smaller variation. The better worst-case performance of LoCUS is important given time limits imposed by battery life.  Additionally, the LoCUS algorithm is able to find the max flux faster than MoBS after initial plume contact, particularly in a perturbed plume simulations.  

For large swarms, the MoBS algorithm makes initial contact with the plume faster than LoCUS on average.  The superior performance of MoBS at finding the plume, and LoCUS of finding the source once in a plume, suggests an approach that combines the best of both algorithms. For large swarms, we may perform the initial search for the plume using the more dispersed golden spoke algorithm used in MoBS.  Then, when contact is made, a LoCUS structured formation can leverage nearby drones to perform gradient descent informed by communication among drones.  Future work can explore the benefits of a fully dispersed set of spokes, with the first drone that contacts the plume calling nearby UAV to join together once the plume is found. Alternatively, sufficiently many UAV could be divided into multiple  small LoCUS sub-swarms to use spoke search to contact the plume and LoCUS enabled gradient descent by each independent sub-swarm once it contacts the plume.

LoCUS relies heavily on its loss recovery model in order to maintain communication between spatially dispersed drones to perform a more robust gradient descent once a plume has been found (see the red lines within the plume in Figure 2).  The loss recovery model allows the swarm to reorganise once a failure has been detected and continue to rely on receiving \ce{CO2} measurements from multiple locations.  In practice, we observed LoCUS successfully locating max flux with failures in a majority of the swarm, even down to a single remaining drone. We also observed that the loss recovery time is so fast that it is dominated by the time to encounter the plume and time to max flux. Thus, the time to recover the swarm formation is worth the superior gradient following performance provided by having spatially dispersed measurements.  In experiments 3 and 4 we show that self-healing is critical to the success of LoCUS gradient following.

In comparison to MoBS, LoCUS is especially vulnerable to the in-plume failures.   This is (ironically) because LoCUS brings the entire swarm into the plume and quickly closer to the source, putting swarm members in jeopardy due to more in-plume failures as the source is approached near the source of volcano efflux.  In contrast, some of the UAV in MoBS spend more time out of the plume, making them less susceptible to in-plume failures. The results in experiment 4 show that even in the worst case for LoCUS, it can leverage the healing algorithm to mitigate this problem to complete the max flux task nearly as often as MoBS.

Being able to reliably and quickly determine the max \ce{CO2} flux with drones that are limited to a maximum \SI{1}{\hour} flight time, with practical swarm sizes for transportation to remote and hazardous regions, and are tolerant of drone loss is critical to the study of volcano behaviour.  With LoCUS, we have demonstrated an algorithm that solves the \ce{CO2} max flux task faster than, and approximately as reliably as a more dispersed approach. 


\section{Acknowledgements}
We thank the UNM Vice President for Research, the Department of Energy’s Kansas City National Security Campus, operated by Honeywell Federal Manufacturing \& Technologies, LLC under contract number DE-NA0002839, a James S. McDonnell
Foundation Complex Systems Scholar Award and DARPA award $\#$FA8650-18-C-6898 for funding. We thank the UNM Center for Advanced Research Computing, supported in part by the National Science Foundation, for providing the high performance computing resources used in this work.

\bibliographystyle{ieeetr}
\bibliography{ref,mendeley}

\end{document}